\documentclass[aip,jcp,twocolumn,superscriptaddress,numerical,reprint]{revtex4}
\usepackage{graphicx}
\usepackage{amssymb}
\usepackage{citesort}

\begin{document}
\title{Subgap Two-Photon States in Polycyclic Aromatic Hydrocarbons: Evidence for Strong Electron Correlations}
\author{Karan Aryanpour}
\email{karana@physics.arizona.edu}
\affiliation{Department of Physics, University of Arizona, Tucson, Arizona 85721, United States}
\author{Adam Roberts}
\affiliation{College of Optical Sciences, University of Arizona, Tucson, Arizona 85721, United States}
\affiliation{U.S. Army Aviation and Missile Research, Development, and Engineering Center, Redstone Arsenal, Huntsville, Alabama 35898, United States}
\author{Arvinder Sandhu}
\affiliation{Department of Physics, University of Arizona, Tucson, Arizona 85721, United States}
\affiliation{College of Optical Sciences, University of Arizona, Tucson, Arizona 85721, United States}
\author{Rajendra Rathore}
\affiliation{Department of Chemistry, Marquette University, Milwaukee, Wisconsin 53201, United States}
\author{Alok Shukla}
\affiliation{Department of Physics, Indian Institute of Technology, Powai, Mumbai - 400076, India}
\author{Sumit Mazumdar$^{*}$}
\affiliation{Department of Physics, University of Arizona, Tucson, Arizona 85721, United States}
\affiliation{College of Optical Sciences, University of Arizona, Tucson, Arizona 85721, United States}
\date{\today}
\begin{abstract}
{Strong electron correlation effects in the photophysics of quasi-one-dimensional
$\pi$-conjugated organic systems such as polyenes, polyacetylenes,
polydiacetylenes, etc., have been extensively studied. Far less is
known on correlation effects in two-dimensional $\pi$-conjugated
systems. Here we present theoretical and experimental evidence for
moderate repulsive electron$-$electron interactions in a number of finite
polycyclic aromatic hydrocarbon molecules with $D_{6h}$ symmetry.
We show that the excited state orderings in these molecules are reversed
relative to that expected within one-electron and mean-field theories.
Our results reflect similarities as well as differences in the role
and magnitude of electron correlation effects in these two-dimensional
molecules compared to those in polyenes.}
\end{abstract}
\maketitle
\section*{$\blacksquare$ INTRODUCTION}
\label{intro}
\begin{figure}
\includegraphics[width=3.0in]{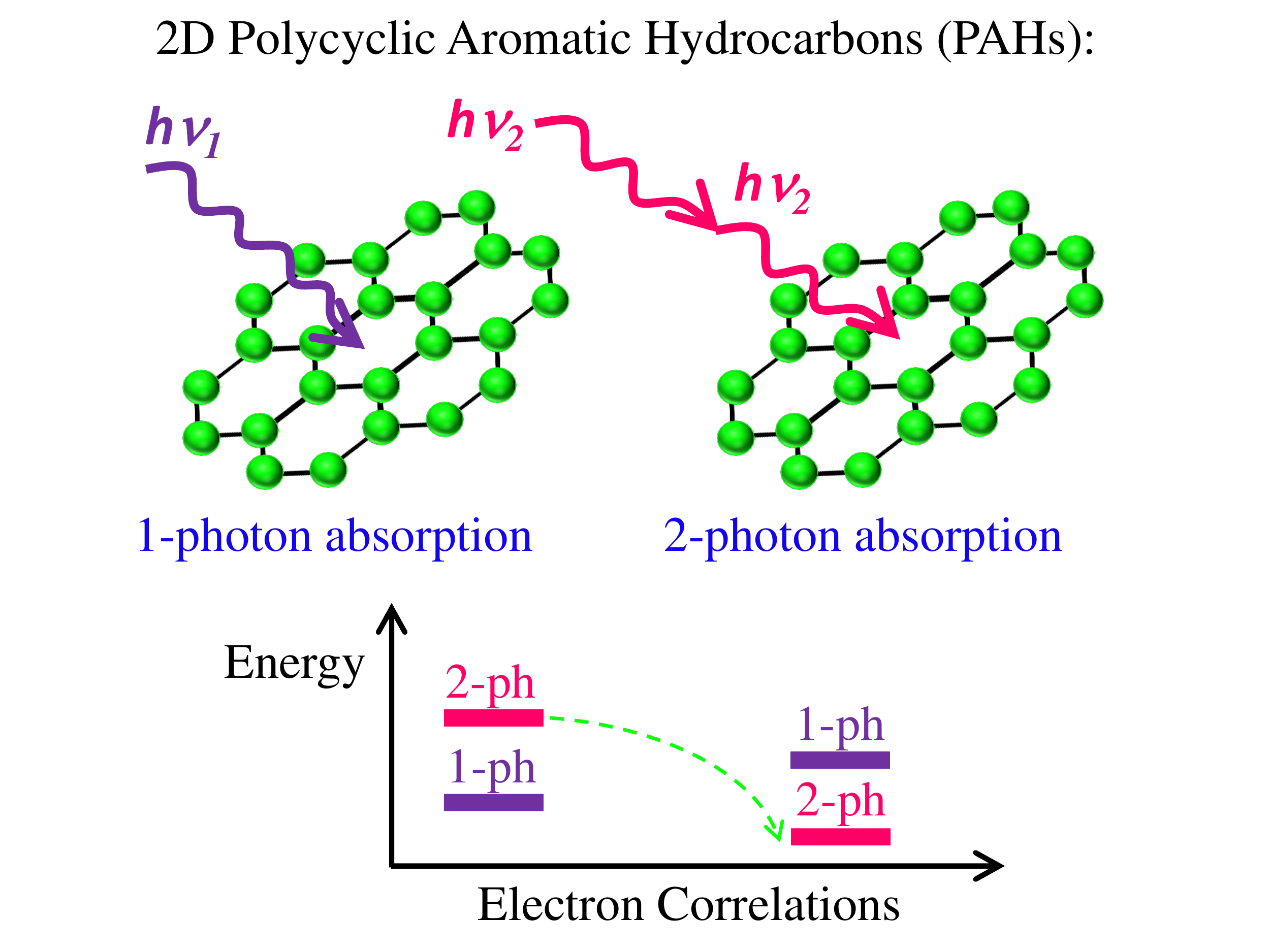}
\\ \center{Table of Contents}
\label{TOC}
\end{figure}
\par Strong electron correlation effects remain a highly relevant topic
in the physics and chemistry of carbon-based materials. Extensive
experimentation has confirmed strong electron correlation effects
in quasi-one-dimensional (quasi-1D) carbon-based systems such as linear
$\pi$-conjugated polymers, single-walled carbon nanotubes, graphene
nanoribbons, etc. Historically, the study of the excited state ordering
in finite polyenes \cite{Hudson82a,Christensen08a,Schulten72a,Ramasesha84a,Tavan87a}
provided the most convincing evidence for strong electron correlation
effects in quasi-1D $\pi$-conjugated systems, and formed the basis
of correlated-electron theories in the long chain polyacetylenes and
polydiacetylenes. \cite{Baeriswyl92a,Barcza13a} The early calculations 
\cite{Schulten72a,Ramasesha84a,Tavan87a} were within the semiempirical 
Pariser$-$Parr$-$Pople (PPP) $\pi$-electron model Hamiltonian,\cite{Pariser53a,Pople53a} 
within which eigenstates of even polyenes with C$_{2h}$ point group symmetry have 
even and odd spatial as well as charge-conjugation symmetries (CCSs), and allowed 
one-photon (two-photon) transitions from the $1^{1}$A$_{g}^{-}$ ground state
are to $^{1}$B$_{u}^{+}$ ($^{1}$A$_{g}^{-}$) states with opposite
(same) symmetries (the superscript $1$ indicates singlet spin state
and the ``plus'' and ``minus'' superscripts refer to CCS). 
The occurrence of the lowest two-photon state, the 2$^{1}$A$_{g}^{-}$,
\textit{below} the one-photon 1$^{1}$B$_{u}^{+}$ in polyenes \cite{Hudson82a,Christensen08a}
is opposite to that expected within the one-electron H\"uckel and mean-field 
Hartree$-$Fock (HF) theories and is a consequence of strong short-range electron$-$electron 
(e$-$e) interactions within the PPP model.\cite{Schulten72a,Ramasesha84a,Tavan87a}
Subsequent to these early model calculations, considerable effort has gone into 
developing more sophisticated quantum chemical correlated {\it ab initio} approaches to 
polyene spectra. Quantum chemical approaches that are now able to reproduce the correct 
one-photon versus two-photon state ordering for short linear polyenes (usually up to octatetraene) 
include the complete active space second order perturbation theory (CASPT2) and 
third-order coupled cluster theory, \cite{Schreiber08a,Silva10a} and the extended algebraic 
diagrammatic construction (ADC(2)-x) method. \cite{Starcke06a,Knippenberg12a,Krauter13a} Correct 
excited state ordering is also found within a density functional theory-based multiple reference 
configuration interaction (DFT-MRCI) approach. \cite{Silva08b} CCS characteristic of the PPP model 
is absent within the latter approaches; however, the violation of CCS is weak. 
\cite{Schmidt12a} Importantly, the quantum chemical approaches do find the double 
excitation character of the 2$^{1}$A$_{g}$. Theoretical works have also found reversed excited state 
ordering in long acenes from pentacene onward, within the PPP model, \cite{Tavan79a,Raghu02a} a 
semi-empirical DFT-MRCI method \cite{Marian08a} and the ADC(2)-x approach. 
\cite{Knippenberg10a} Acenes can be considered as coupled polyene chains and are also quasi-1D. 
\par
\begin{figure}
\includegraphics[width=2.0in]{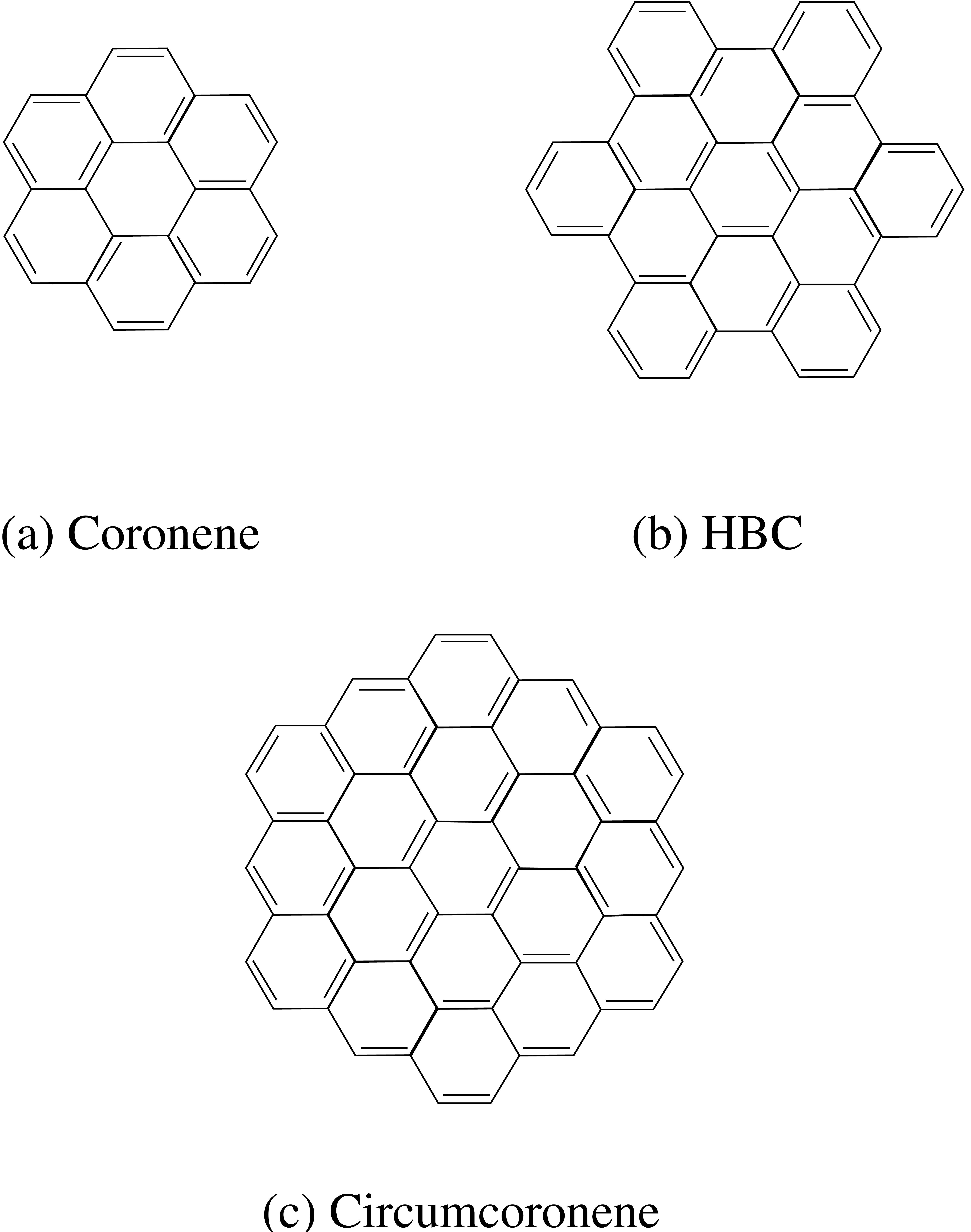}
\caption{PAH molecules studied in this work: (a) coronene (C$_{24}$H$_{12}$)
(b) hexa-peri-hexabenzocoronene, hereafter HBC (C$_{42}$H$_{18}$),
and (d) circumcoronene (C$_{54}$H$_{18}$).} 
\label{f1}
\end{figure}

 In contrast to the quasi-1D systems, there have been few, if any,
systematic studies of electron correlation effects in two-dimensional
(2D) molecules. In this work we take the first step toward addressing
electron correlation effects in 2D molecules by conducting a systematic
study of excited states ordering in polycyclic aromatic hydrocarbons
(PAHs). By performing meticulous experimental and computational analyses
of the one- and two-photon excited states for the three PAH molecules
shown in Figure~\ref{f1}a$-$c, we find reversed excited state ordering
within these 2D molecules in clear contradiction to the description
of one-electron theory. These correlation effects are found to be
overall weaker compared to quasi-1D carbon-based systems, due to the
larger one-electron bandwidth in 2D. Possible extension of our study
to larger molecules can also shed light on electron correlation effects
in graphene as a genuine 2D carbon-based system in the thermodynamic
limit. Within the H\"uckel tight-binding $\pi$-electron theory graphene
is a semimetal, with linear energy versus momentum relationship near
the Dirac point where the filled valence and empty conduction bands
meet. Although many experimental observations appear to be in agreement
with this description, \cite{Geim07a,Neto09a} the neglect of the
repulsive e$-$e interaction between the $\pi$-electrons is increasingly
being questioned. \cite{Elias11a,Kotov12a} 
\section*{$\blacksquare$ THEORETICAL MODEL AND METHODS}
\label{model}
\subsection*{Pariser$-$Parr$-$Pople Model}
\par Quantum chemical approaches can correctly describe low-lying two-photon states of double excitation character only for relatively small molecules. \cite{Schreiber08a,Silva10a,Starcke06a,Knippenberg12a,Krauter13a,Silva08b,Schmidt12a,Tavan79a,Marian08a,Knippenberg10a}
The two larger PAH molecules of interest here, namely, HBC and circumcoronene,
lie outside the scope of these calculations. Moreover, while we are
principally interested in the relative energies of the lowest one-
and two-photon states, we will in the following compare experimental
and calculated two-photon absorption (TPA) spectra over a broad frequency
range, which requires that we are able to calculate the energies and
wave functions of the higher energy two-photon states. These requirements
restrict us to perform our calculations within the semiempirical PPP Hamiltonian,
\cite{Pariser53a,Pople53a} which is written as
\begin{eqnarray}
 H_{PPP}=-t\sum_{\langle ij \rangle\sigma}
(c_{i\sigma}^\dagger c_{j\sigma}+c_{j\sigma}^\dagger c_{i\sigma}) + U\sum_{i} n_{i\uparrow} n_{i\downarrow} 
 + \nonumber \\ \sum_{i<j} V_{ij} (n_{i}-1)(n_{j}-1)\,,\hspace{4.0cm}
\label{PPP_Ham}
\end{eqnarray}
where $c_{i\sigma}^{\dagger}$ creates a $\pi$-electron of spin $\sigma$
on carbon atom $i$, $n_{i\sigma}=c_{i\sigma}^{\dagger}c^{}_{i\sigma}$
is the number of electrons of spin $\sigma$ on atom $i$, and $n_{i}=\sum_{\sigma}n_{i\sigma}$.
The one-electron hopping integral $t$ is between nearest neighbor
carbon atoms $i$ and $j$, $U$ is the Hubbard repulsion between
two electrons occupying the same atomic $p_{z}$ orbital, and $V_{ij}$
is the long-range intersite Coulomb interaction.
We choose standard $t=2.4$ eV, \cite{Ramasesha84a,Tavan87a,Baeriswyl92a}
and obtain $V_{ij}$ from the parametrization, \cite{Chandross97a}
$V_{ij}=U/\kappa\sqrt{1+0.6117R_{ij}^{2}}$, where $R_{ij}$ is the
distance in $\mathring{\textrm{A}}$ between carbon atoms $i$ and
$j$ and $\kappa$ is an effective dielectric constant. We fix carbon$-$carbon
(C$-$C) bond lengths to $1.4$ $\mathring{\textrm{A}}$ and all bond
angles to 120$^{\circ}$. The actual bond lengths and angles may deviate
slightly from these mean values, but given the aromatic nature of
the molecules, we expect these deviations and their effects on the
energies of the one- and two-photon states to be small. This is shown
explicitly for coronene (see Appendix),
the only molecule for which the experimental bond lengths are available
in the literature. We choose $U=8$ eV and $\kappa=2$ based on the
excellent fits obtained previously with these parameters for the excited
states of poly$-$paraphenylenevinylene, \cite{Chandross97a} polyacenes,
\cite{Sony07a} and single-walled carbon nanotubes. \cite{Wang06a} 
\par The molecules in Figure~\ref{f1} belong to the $D_{6h}$ point group and
possess inversion, reflection, 6-fold rotational, and CC symmetries. 
Since $D_{6h}$ is not Abelian, in the following we adopt
the Abelian $D_{2h}$ point group symmetry to partition excited states
into distinct one- and two-photon classes. The ground state is in
the $^{1}$A$_{g}^{-}$ subspace for all the molecules of Figure~\ref{f1}a$-$c.
Dipole selection rules dictate that one-photon (linear) absorption (TPA)
occur only to $^{1}$B$_{2u}^{+}$ and $^{1}$B$_{3u}^{+}$ ($^{1}$A$_{g}^{-}$
and $^{1}$B$_{1g}^{-}$) states with spatial and CC symmetries opposite to (the same as) 
that of the ground state. In addition PAHs also have $^{1}$B$_{2u}^{-}$ and $^{1}$B$_{3u}^{-}$ states
forbidden in both one- and two-photon processes. \cite{Sony07a,Wang06a} 
\subsection*{Multiple-Reference Configuration Interaction Approach}
\par Our calculations were done using the multiple reference singles and
doubles configuration interaction (MRSDCI) \cite{Tavan87a}. The MRSDCI
retains the dominant single, double, triple, and quadruple excitations
for each targeted state \cite{Tavan87a}, with the number of configurations
systematically increased until numerical convergence is reached.
 While excitonic effects due to the correlation between
a single electron$-$hole pair in optically allowed states have been
treated very successfully for graphene \cite{Yang09a} and other carbon
nanostructures \cite{Spataru04a,Chang04a,Prezzi08a} within the GW-Bethe
Salpeter technique, application of this first-principles approach
to investigate the lowest two-photon eigenstates has not yet been
possible. For example, the lowest two-photon states that occur below
the optical gap in trans-polyacetylene \cite{Baeriswyl92a} is missed
within the GW-Bethe Salpeter approach, \cite{Rohlfing99a} even though
the latter gives accurate description of the optical exciton. Thus,
the inclusion of CI with multiple electron$-$hole excitations is a necessary
requirement for obtaining two-photon states below the lowest one-photon
optical state. 
\cite{Hudson82a,Schulten72a,Ramasesha84a,Tavan87a,Baeriswyl92a,Schreiber08a,Silva10a,Starcke06a,Knippenberg12a,Krauter13a,Silva08b,
Schmidt12a,Tavan79a,Marian08a,Knippenberg10a}
The semiempirical approach taken by us cannot be easily extended
to the thermodynamic limit (except in one dimension, where the density
matrix renormalization group approach can be used \cite{Ramasesha00a,Bursill02a,Barcza13a}),
but this disadvantage is offset by the ability to include high order
correlation effects, involving triply and quadruply excited configurations
with respect to the restricted Hartree$-$Fock (RHF) reference state. Next, we describe the procedure
involved in performing various levels of CI calculations presented
in this work.
\par The initial step of our CI studies involves determination of the self-consistent
RHF solutions of the PPP Hamiltonian (see eq~\ref{PPP_Ham}). 
The basis functions of the many-body calculations
are configurations with all possible electron occupancies of the HF molecular orbitals (MOs). 
A full CI calculation involves construction of the Hamiltonian
matrix with all possible excitations from the HF ground state, and
this is the procedure we have followed in the text for octatetraene
and decapentaene. For select few cases, quadruple-CI (QCI), which
incorporates all excitations up to quadruples from the HF ground state,
is also possible. With a large number of electrons, as in the present
case, one then encounters the following problem. On one hand,
straightforward high-order CI such as the QCI is beyond today's
computational capability. On the other hand, lower order CI such as double-CI
does not give the correct excited state ordering for large molecules.
It is precisely for such large systems that the MRSDCI is used. \cite{Buenker74,Tavan86}
We have previously performed calculations as large as the ones reported
here for quasi-1D molecules and oligomers. \cite{Ghosh00,Shukla02a,Shukla03a,Shukla04a,Shukla04b,
Sony05a,Aryanpour11a} 
\par The MRSDCI is performed in two steps. At the first step, a few ($N_{ref}$)
singly and doubly excited reference configurations that best describe
the excited states in the targeted symmetry subspace are selected
on the basis of a trial double-CI calculation. The second step involves
the MRSDCI calculation, in which the Hamiltonian matrix consists of
single and double excitations with respect to the original $N_{ref}$
configurations themselves. The total number of configurations $N_{total}$
therefore includes also the subset of the dominant triple and quadruple
excitations. The new set of $N_{total}$ configurations are now examined,
single and double excitations from the HF ground states that contribute
significantly to the eigenstates within the targeted symmetry subspace
are identified, and these are included to augment the new reference
space, following which the MRSDCI is performed again. This iterative
approach is continued, with updating of $N_{ref}$ and $N_{total}$
at every step, until convergence to some tolerance determined by the
size of the system is reached. 
\begin{table*} 
\begin{tabular}{l|r}
{symmetry} & {coronene\hspace{1.9cm}HBC\hspace{1.5cm}circumcoronene} \\ 
& {$N_{ref}$\hspace{0.7cm}$N_{total}$\hspace{0.8cm}$N_{ref}$\hspace{0.7cm}$N_{total}$\hspace{0.7cm}$N_{ref}$\hspace{0.7cm}$N_{total}$}\\  
\hline \\
{$^1$B$_{2u}$:} & {$215$\hspace{0.45cm}$1082466$\hspace{0.9cm}$102$\hspace{0.4cm}$2227463$\hspace{1.1cm}$77$\hspace{0.55cm}$3645309$}  \\ 
{$^1$B$_{3u}$:} & {$202$\hspace{0.45cm}$972754$\hspace{1.15cm}$88$\hspace{0.5cm}$1719854$\hspace{1.1cm}$70$\hspace{0.55cm}$3133234$}  \\ 
{$^1$A$_{g}$:} & {$1^{b}$\hspace{0.45cm}$2045687$\hspace{0.95cm}$184$\hspace{0.4cm}$3371103$\hspace{1.1cm}$100$\hspace{0.4cm}$3864837$}  \\
{$^1$B$_{1g}$:} & {$268$\hspace{0.45cm}$1162244$\hspace{0.95cm}$168$\hspace{0.4cm}$3167504$\hspace{1.1cm}$148$\hspace{0.4cm}$3745386$} \\
\end{tabular} 
\caption{Number of Final Reference Configurations, $N_{ref}$, and the Dimension
of the Hamiltonian Matrix, $N_{total}$, for the PAH Molecules of Figure~\ref{f1}a$-$c 
for the $D_{2h}$ Symmetry Subspaces Relevant
to Linear and Nonlinear Optics$^{a}$\protect \\ \\
\protect 
 {\footnotesize{$^{a}$The final MRSDCI wavefunctions contained
basis functions with coefficients as small as 0.04, 0.04 and 0.06
for coronene, HBC and circumcoronene, respectively. $^{b}$QCI}} }
\label{t1} 
\end{table*}
\par In Table~\ref{t1} we have given the final $N_{ref}$ and $N_{total}$
reached in our MRSDCI calculations for all four $D_{2h}$ subspaces
relevant for the calculations of nonlinear absorptions in coronene,
HBC and circumcoronene (for the $^{1}$A$_{g}$ subspace of coronene
the calculation was done using QCI). To the best of our knowledge,
these are the largest CI calculations performed to date for the excited
states of these molecules. 
\subsection*{Calculation of the Two-Photon Absorption Spectrum}
The TPA is calculated from the imaginary component
of the third-order nonlinear optical susceptibility $\chi^{(3)}(\omega,\omega,\omega,-\omega)$,
the resonant contribution $\chi^{(3)}(TPA)$ of which is given by
the sum-overstate expression, \cite{Boyd92a} 
\begin{eqnarray}
\chi_{ijkl}^{(3)}(TPA)=\frac{N}{6\hbar^{3}}\Bigg({\sum_{mnp}}^{'}\big[\mu_{gn}^{i}
\mu_{nm}^{l}\mu_{mp}^{k}\mu_{pg}^{j}+\hspace{1.0cm}\nonumber \\\mu_{gn}^{i}\mu_{nm}^{l}\mu_{mp}^{j}\mu_{pg}^{k}+\mu_{gn}^{l}
\mu_{nm}^{i}\mu_{mp}^{k}\mu_{pg}^{j}+\mu_{gn}^{l}\mu_{nm}^{i}\mu_{mp}^{j}\mu_{pg}^{k}\big]\times\hspace{0.7cm}\nonumber \\
\frac{1}{(\omega_{ng}-\omega)(\omega_{mg}-2\omega-i\delta)(\omega_{pg}-\omega)}-
\sum_{np}\big[\mu_{gn}^{i}\mu_{ng}^{j}\mu_{gp}^{l}\mu_{pg}^{k}\hspace{0.7cm}\nonumber \\+\mu_{gn}^{i}\mu_{ng}^{k}\mu_{gp}^{l}\mu_{pg}^{j}+
\mu_{gn}^{l}\mu_{ng}^{k}\mu_{gp}^{i}\mu_{pg}^{j}+\mu_{gn}^{l}\mu_{ng}^{j}\mu_{gp}^{i}
\mu_{pg}^{k}\big]\times\hspace{0.7cm}\nonumber \\\frac{1}{(\omega_{ng}-\omega)(\omega_{pg}-\omega)(\omega_{pg}-\omega)}\Bigg)\,,\hspace{1.0cm}
\label{TPA}
\end{eqnarray}
where $N$ is the number of molecules, $g$ is the ground state, $n$
and $p$ are virtual one-photon states, and $m$ is two-photon states. $\delta$ 
is the line width of the 
spectrum when $\omega\approx\omega_{mg}/2$ and was set to $0.03$ eV in all 
our $\chi^{(3)}(TPA)$ calculations for various PAH molecules in this work.
Here $\mu_{gn}^{i}$ is the $i$th component of the matrix element
of the transition dipole operator between states $g$ and $n$. Other
components of the transition dipole matrix elements are defined similarly,
and the prime over the first summation indicates $m\neq g$. With
$D_{2h}$ symmetry, TPAs are to $^{1}$A$_{g}^{-}$
and $^{1}$B$_{1g}^{-}$ states, which are nondegenerate when the Coulomb
interactions are nonzero. The matrix element of the $x$-component
of the transition dipole operator is nonzero only between the $^{1}$A$_{g}^{-}$
and the $^{1}$B$_{3u}^{+}$ states. As a consequence, two-photon
resonances in the $1111$, $2222$, $1221$, and $2112$ components of $\chi^{(3)}(TPA)$ in eq \ref{TPA}
are all to $^{1}$A$_{g}^{-}$ states only. Similarly, the matrix element
of the $y$-component of the transition dipole operator is nonzero
between $^{1}$B$_{3u}^{+}$ and $^{1}$B$_{1g}^{-}$ states only.
Hence, two-photon resonances in the $1212$, $2121$, $1122$, and $2211$ components of $\chi^{(3)}(TPA)$
are all to $^{1}$B$_{1g}^{-}$ states. In eq \ref{TPA} above, the intermediate
one-photon virtual state of $^{1}$B$_{1u}^{+}$ and two-photon states of $^{1}$B$_{2g}^{-}$, and $^{1}$B$_{3g}^{-}$ do not contribute 
because they all involve matrix elements of the $z$ component (out of plane) of the dipole operator; due to the planar
 structure of PAH molecules studied in this work, these dipole 
 transition matrix elements are vanishingly small. The experimental solution-phase PA spectrum is 
 compared against the orientational average of the gas-phase TPA susceptibilities
 \cite{Heflin88a},
\begin{eqnarray}
\chi_{avg}^{(3)}(TPA)=\frac{1}{5}\Bigg[\sum_{i}\chi_{iiii}^{(3)}(TPA)+\frac{1}{3}\Big(\sum_{i\neq j}\big(\chi_{iijj}^{(3)}(TPA)\nonumber \\ +
\chi_{ijij}^{(3)}(TPA)+\chi_{ijji}^{(3)}(TPA)\Big)\Bigg]\,.\hspace{1.0cm}
\label{TPA-AVG}
\end{eqnarray}
\section*{$\blacksquare$ RESULTS: THEORY VERSUS EXPERIMENT}
\label{results} 
\par{\bf\noindent H\"uckel Calculations.} Before presenting the results of the
correlated-electron calculations, we discuss the results of calculations
based on the H\"uckel model which is the $U=V_{ij}=0$ limit of the
PPP model. The goal here is to demonstrate that the H\"uckel model fails
\textit{qualitatively} to give the excited state ordering that is
observed experimentally (see below). In Figure~\ref{f2}a$-$c
we show the frontier MOs for coronene, HBC and
circumcoronene, respectively, where we have indicated the highest
occupied and lowest unoccupied MOs (HOMO and LUMO) and the lowest
one and two-photon allowed one electron-one hole (1e$-$1h) transitions
(1e$-$1h two-photon transitions that originate from MOs below the HOMO
and are related to the ones shown in the figure by CCS are not shown). 
Note that all energies are in units of $|t|$
and dimensionless here. Two electron-two hole (2e$-$2h) two-photon states
lie outside the region of experimental interest and are hence not
shown. Optically bright and dark states of B$_{u}$ symmetry are degenerate
in the H\"uckel limit. The two-photon $^1$A$_{g}^{-}$ and $^1$B$_{1g}^{-}$
states are also degenerate. 
\begin{figure*}
\includegraphics[width=6.2in]{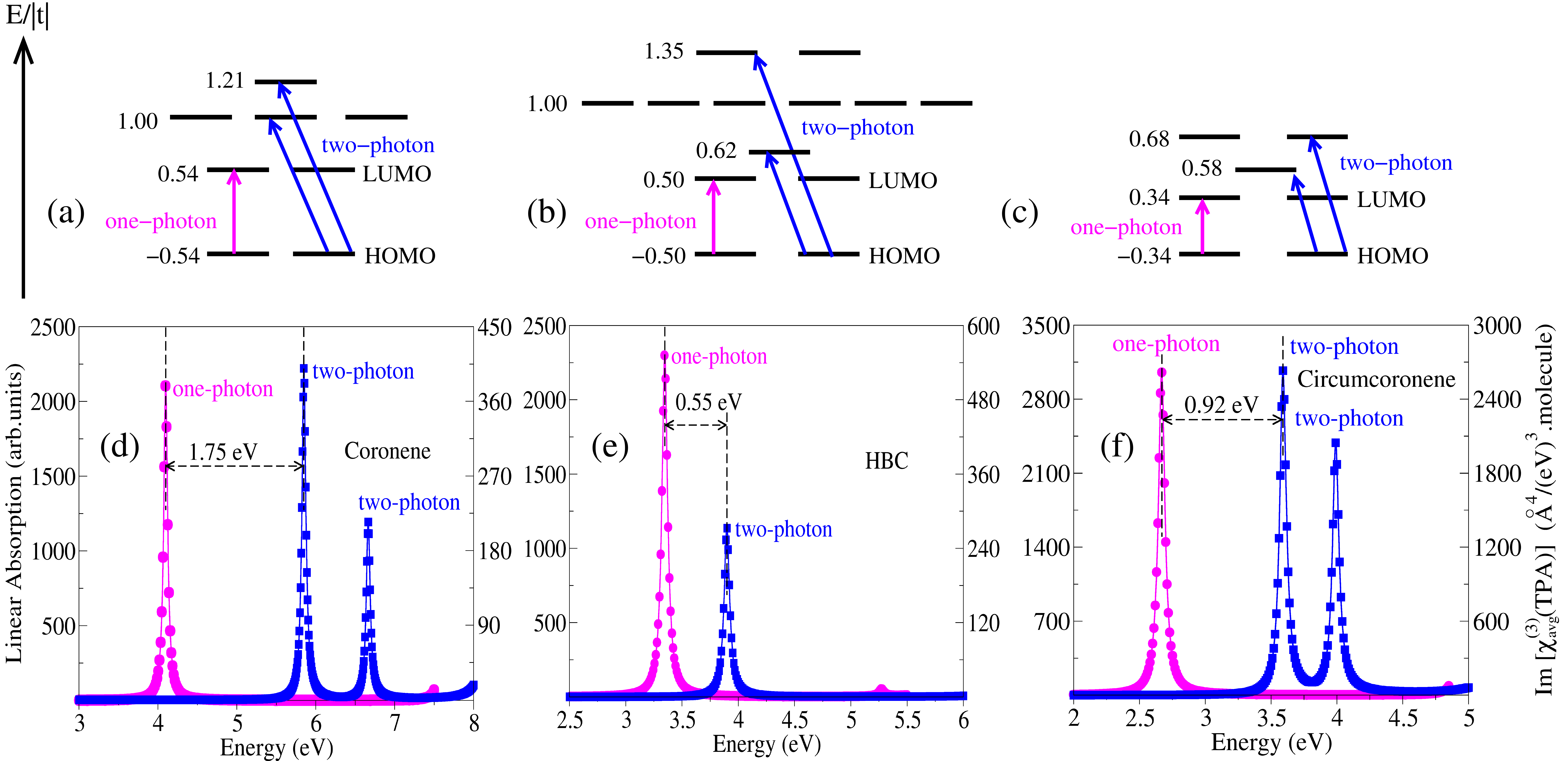}
\caption{Frontier MOs of (a) coronene, (b) HBC, and (c) circumcoronene, with the
lowest one-photon and two-photon allowed 1e$-$1h excitations indicated
by arrows in the noninteracting H\"uckel limit. Panels (d)$-$(f) present the calculated 
linear and TPA spectra corresponding to molecules of panels (a)$-$(c), respectively. Two-photon
transitions shown are at twice the fundamental frequencies, for comparison
to the one-photon transitions. The $t_{ij}$ used in the calculations
are $3.8$ eV for both coronene and circumcoronene and $3.6$ eV for HBC.
These values were arrived at by fitting the peak energies of the strong
optical absorptions in the experimental linear spectra of coronene
and HBC, and the calculated many-body optical gap in circumcoronene
against the H\"uckel theory (see text). $\delta$ in eq \ref{TPA} is $0.03$ 
eV for all the TPA spectra of panels (d)$-$(f).} 
\label{f2}
\end{figure*} 
\par The hopping integral $t$ determines all orbital energies and energy
gaps within the H\"uckel model. It is thus possible to fit the allowed
optical gap for each molecule by choosing an appropriate $t$, although
this may need slightly different $t$ values for different molecules.
Furthermore, in the absence of correlation contribution to the optical
gap, it is only natural that the absolute value of $|t|$ needed to fit the optical
gap within the H\"uckel model will be larger than in the PPP model.
Fitting the peak energy $4.1$ eV of the strong optical absorption in
coronene within the H\"uckel theory requires $t=3.8$ eV. The calculated
linear absorption spectrum with the larger $|t|$ is shown in Figure~\ref{f2}d. 
Figure~\ref{f2}d shows our plot of Im{[}$\chi_{avg}^{(3)}(TPA)${]} 
(see eq \ref{TPA-AVG}) for coronene, which was evaluated using the exact 
H\"uckel wave functions and the sum-overstates procedure of eq \ref{TPA}. 
 The lowest two-photon state in Figure~\ref{f2}d is $1.75$ eV above the optical 
one-photon state in coronene. In discussions of excited state ordering, 
the focus is usually on energetics alone. We emphasize that the density 
of two-photon states and the intensities of the TPAs also reveal correlation 
effects, \cite{McWilliams91a,FGuo94a} and note that in Figure~\ref{f2}d the TPA 
is most intense at the lowest energy. 
\par Figure~\ref{f2}e shows the H\"uckel plots of linear absorption and TPA
in HBC, again with the calculated one-photon gap matching the experimental
gap. The required $t$ is now $3.6$ eV, and the lowest two-photon
state is $0.55$ eV above the one-photon state. We chose the same $t=3.8$ eV 
in our calculations for circumcoronene as in coronene (H\"uckel
calculation of circumcoronene with this $t$ reproduces the optical
gap obtained with PPP-MRSDCI and $t=2.4$ eV, as given in Table~\ref{t2}
below). The calculated H\"uckel linear and TPA spectra
for circumcoronene are shown in Figure~\ref{f2}f. Once again, the lowest
two-photon state is considerably higher than the one-photon state,
and as in coronene the higher energy two-photon state has a weaker
intensity than the lower one. The similarities between the spectra
of coronene and circumcoronene, and their differences with the spectra
of HBC (smaller energy difference between lowest two-photon and one-photon
states, much larger energy difference between the two-photon states)
arise from the different characters of the edges, namely, zigzag versus
armchair. \cite{Guclu10a} 
\par To summarize, the energy gaps to the lowest one-photon optically allowed
states, observed experimentally, can always be reproduced within the
H\"uckel model by choosing an appropriate one-electron hopping integral
$t$ (usually larger than realistic). However, as seen in the ``TPA Measurement'' subsection below,
 the same procedure, when extended to two-photon states, not only fails to give correct
fits to experimentally observed energies of two-photon states, but
actually predicts incorrect excited state ordering. 
\begin{table*}
\begin{tabular}{l|r}
{ molecule} & { S$_2$ (eV)\hspace{0.6cm} 2$^1$A$_g^{-}$ (eV)\hspace{1.8cm}T$_1$ (eV)\hspace{1.0cm}$\xi_{1{\mathrm e}-1{\mathrm h}}$} \\ \\
\hline \\
{octatetraene} & { $4.50~(4.40^{b})$\hspace{0.4cm}$3.42~(3.59^{b})$\hspace{1.3cm}$1.65~(1.73^{c})$\hspace{1.0cm}$0.33$}  \\ \\ 
{decapentaene} & { $4.13~(4.02^{b})$\hspace{0.4cm}$3.06~(3.10^{b})$\hspace{1.8cm}$1.52$\hspace{1.6cm}$0.30$}  \\ \\ 
{coronene} & { $4.14$\hspace{1.5cm}$3.96$\hspace{1.2cm}$2.38~([2.37-2.40]^{d})$\hspace{0.65cm}$0.63$}  \\ \\ 
{HBC} & {$3.37$\hspace{1.5cm}$3.30$\hspace{2.5cm}$2.07$\hspace{1.6cm}$0.71$}  \\ \\ 
{circumcoronene} & {$2.67$\hspace{1.5cm}$2.75$\hspace{2.5cm}$1.50$\hspace{1.6cm}$0.60$}  \\ \\ 
\end{tabular} 
\caption{The Calculated Energies of the One-Photon Optical State S$_{2}$,
the Lowest Two-Photon State 2$^{1}$A$_{g}^{-}$, and the Lowest Triplet
State T$_{1}$ for Linear Polyenes and the Polycyclic Molecules of
Fig.~\ref{f1}$^{a}$  \protect \\ \\
 $^{a}$The numbers in parentheses are experimental quantities. The last column gives the fractional contribution by 1e$-$1h HF excitations to the 2$^{1}$A$_{g}^{-}$. $^{b}$Table II, ref~\onlinecite{Hudson82a}.$^{c}$Reference~\onlinecite{Allan84a}. $^{d}$Reference~\onlinecite{Abouaf09a}. \protect \\
}
\label{t2} 
\end{table*}
\par
\begin{figure*}
\includegraphics[width=6.2in]{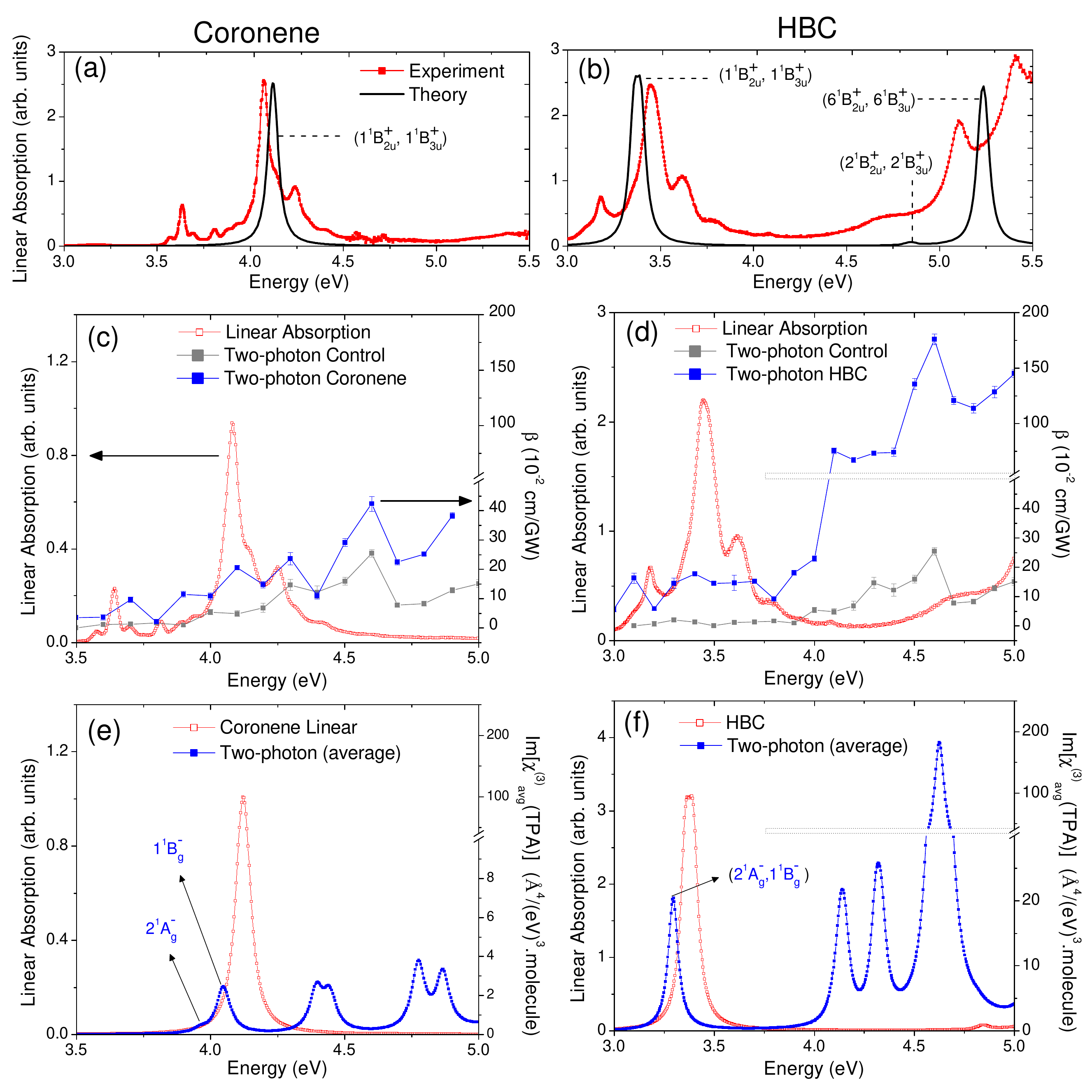}
\caption{(a) Experimental (red) and calculated (black) optical absorption spectra
of coronene; calculations are within the PPP model. The weak absorption
at $\sim$ 3.5 eV, missed in the calculation, is to a state that is
optically dark within the purely electronic PPP Hamiltonian with nearest-neighbor
electron hopping. (b) Same as in (a) for HBC. The lowest energy weak
absorption is again to a dark state. (c) Experimental linear (red)
and TPA (blue) spectra of coronene. The gray curve
gives the TPA due to the solvent. Notice the significant
TPA of coronene below the linear absorption edge.
(d) Same as (c) for HBC. (e) Calculated linear absorption (red) and average TPA Im$[\chi_{avg}^{(3)}(TPA)]$
(blue) for coronene. Two-photon resonances are to both $^1$A$_{g}$ and $^1$B$_{1g}$ states (arrows) which are nondegenerate
within the PPP model. (f) Same as in (e), for HBC. $\delta$ in eq \ref{TPA} is $0.03$ 
eV for the TPA spectra of panels (e) and (f).} 
\label{f3}
\end{figure*}

{\bf\noindent Experimental Linear Absorption versus PPP-MRSDCI Calculations.} 
The linear absorption measurements were
performed with a UV/visible spectrophotometer
(Varian Inc.) in dilute ($\sim$0.1 mM) solutions of Coronene and HBC in dichloroethane. 
The coronene sample was obtained from Sigma-Aldrich Chemical Co. and HBC was prepared 
using the method outlined in ref~\onlinecite{Rathore03a}. We observed that the main features 
of absorption spectra do not change for concentrations between 0.01 and 1 mM, indicating that 
aggregation effects, if any, are minimal.
Figure~\ref{f3}a,b present the experimental solution phase linear absorption
 spectra of coronene and HBC,
superimposed on our calculated PPP-MRSDCI absorption spectra. All
calculated PPP spectra, here and in the following, are with uniform
$t=2.4$ eV.  We find excellent match between the strong experimental absorption band and
the calculated absorption band corresponding to transitions to the
degenerate 1$^{1}$B$_{2u}^{+}$ and 1$^{1}$B$_{3u}^{+}$ states
in both coronene and HBC. The weak splitting in the strong experimental
absorption band, seen in both cases, confirms the existence of two
underlying transitions that have lost their perfect degeneracy in
the real systems. The much weaker absorptions (at $\sim3.55$ eV in
coronene and $\sim3.2$ eV in HBC), missing in the calculated absorption
spectra, are to the ``forbidden'' 1$^{1}$B$_{2u}^{-}$ and 1$^{1}$B$_{3u}^{-}$
states. Loss of degeneracy of the allowed transitions, and nonzero
oscillator strengths of the ``forbidden'' transitions can be due
to the couplings of electrons with vibronic modes as well as weak
second neighbor electron hopping that are absent in our Hamiltonian.
Neither of these interactions affect the relative ordering of the
one- and two-photon states. 
\par{\bf\noindent TPA Measurement.} We conducted experimental measurements of TPA
 in $\lesssim$ 1 mM solutions of coronene and HBC using dichloroethane
as a solvent. High-intensity, 100 fs laser pulses with photon
energies between 1.5$-$3 eV were obtained from titanium-sapphire
pumped optical parametric amplifier (OPA) and used to measure the
TPAs to states at energies 3.0$-$6.0 eV. The coronene
and HBC solutions in 1 mm UV-quartz cuvettes were used for open-aperture
``z-scan'' measurement that allows extraction of TPA
coefficient. \cite{SheikBahae90a} We employed a modified differential
z-scan approach similar to the one discussed in ref~\onlinecite{Menard07a}
to enhance the sensitivity of our TPA measurements.
The z-scan measurements with the solvent alone were used as the ``control''
for the data reported here. 
\par In Figure~\ref{f3}c we show the experimental TPA spectrum
of coronene, where we have included the linear absorption for
comparison. The TPA spectrum of Figure~\ref{f3}c is different
from the H\"uckel TPA spectrum for coronene in Figure~\ref{f2}d in
three distinct ways: (i) resonances to two-photon states \textit{below}
the allowed one-photon states are observed, (ii) the experimental
spectrum shows many more structures, and (iii) the intensities of
the TPAs are higher at higher energy. The TPA 
spectrum of HBC, shown in Figure~\ref{f3}d is very similar to
that of coronene, with once again nonzero TPA below
the strong one-photon allowed linear absorption. The strengths of
the TPAs, both below the optical gap and at higher
energies, are significantly higher for HBC with larger molecular size.
A similar increase in the intensities of the TPAs
for small chain lengths is also observed in polyenes. \cite{FGuo94a}
We again note the apparent high density of two-photon states in the
energy region 4.0$-$5.0 eV, in contradiction to the prediction of
the H\"uckel theory (see Figure~\ref{f2}e). The energy-dependence of the intensities
for the TPAs in HBC is also similar to that in coronene. 
\par{\bf\noindent PPP-MRSDCI Calculations of TPA.} Figure~\ref{f3}e,f presents the calculated TPA spectra 
(eq~\ref{TPA-AVG}). Dipole matrix elements were evaluated using the PPP-MRSDCI many-body
wave functions. The calculated  linear spectra have been
included in the figures for comparison. The $\chi_{avg}^{(3)}(TPA)$ spectrum has resonances to both 
$^{1}$A$_{g}^{-}$ and $^{1}$B$_{1g}^{-}$ states (see arrows in Figure~\ref{f3}e,f).
\par In agreement with experiments, the calculated lowest two-photon state,
the 2$^{1}$A$_{g}^{-}$, occurs below the optical 1$^{1}$B$_{2u}^{+}$
and 1$^{1}$B$_{3u}^{+}$ states in coronene (by $0.18$ eV) as well
as in HBC (by $0.08$ eV). Comparing Figure~\ref{f3}d,f we find excellent
matches between the energy locations of the experimental and calculated
two-photon resonances in HBC. The fit is less impressive for coronene.
The high densities of two-photon states observed in both coronene
and HBC agree with the occurrence of many close lying $^{1}$A$_{g}^{-}$
and $^{1}$B$_{1g}^{-}$ states in the same energy range within the
PPP model. The calculations also reproduce the intensity profile of
the experimental TPAs, with the intensity at higher
energies larger than that at lower energies. 
\par The calculated lowest two-photon state is slightly above the lowest
one-photon state in circumcoronene (see Table~\ref{t2} below), with, however,
an energy separation ($0.08$ eV) much smaller than expected within
one-electron theory. The almost ``normal'' excited state ordering
in circumcoronene is likely due to a weak breakdown of the MRSDCI
approximation, which becomes less appropriate with increasing size.\cite{Schmidt12a}
With $54$ C-atoms, circumcoronene is more than twice the size of coronene
and is near the memory limit of MRSDCI, {\it viz.}, $60$ atoms. 
\section*{$\blacksquare$ DISCUSSION AND CONCLUSIONS}
\label{discussion} 
\par Correlation effects in 2D, though substantial, are weaker than in
linear chains. Table~\ref{t2} summarizes the calculated PPP energies of
the lowest one- and two-photon spin singlet states, the lowest triplet
state (hereafter T$_{1}$), and the wave function characteristics of
the lowest two-photon state for 1,3,5,7-octatetraene, 1,3,5,7,9-decapentaene,
and the three molecules we have investigated here. The polyene full
CI calculations were done with hopping integrals $t=2.6$ and $2.2$
eV for the double and single bonds, respectively.\cite{Chandross97a}
We have also included in Table~\ref{t2} all experimental energy values
that were available to us from literature. We note the excellent agreements
between all theoretical and experimental energies for the two polyenes.
Equally remarkable agreement is found between the calculated
and experimental triplet energies of coronene. The 2$^{1}$A$_{g}^{-}$
in the polyenes is known to be a quantum-entangled ``two-triplet''
state \cite{Ramasesha84a,Tavan87a} whose energy is nearly twice that
of T$_{1}$. By contrast, 2$^{1}$A$_{g}^{-}$ in the PAH molecules in all 
cases is less than twice the energy of their T$_{1}$. The low energy 
of the 2$^{1}$A$_{g}^{-}$ in polyenes,
relative to the 1$^{1}$B$_{u}^{+}$, is best understood within valence bond (VB) theory. \cite{Ramasesha84a,Tavan87a}
For strong Hubbard $U$, the ground state within VB theory is predominantly covalent, {\it i.e.}, its wave function
is dominated by VB diagrams in which all carbon atom $p_z$-orbitals are singly occupied. Optical absorption within tight-binding Hamiltonians necessarily requires charge-transfer between carbon atoms,\cite{Ramasesha84a,Tavan87a} which then implies that the 1$^{1}$B$_{u}^{+}$ is dominated by ionic VB diagrams with at least one pair of C$^+$C$^-$ ion pair. Two-photon states are reached by a {\it second} charge-transfer excitation from a virtual ionic state, which 
can create yet another C$^+$C$^-$ ion pair, or annihilate the original ion pair, creating an excited state that is also covalent, but orthogonal to the
ground state. \cite{Ramasesha84a,Tavan87a} For nonzero Hubbard $U$, the lowest ionic states are necessarily higher in energy
than the lowest covalent states. There is a one-to-one correspondence between this qualitative VB picture 
and the CI approach within the MO basis, in that the 1$^{1}$B$_{u}^{+}$, obtained by a single charge-transfer from the ground state
in configuration space, is an 1e$-$1h excitation, while the 2$^{1}$A$_{g}^{-}$ has strong contributions from 2e$-$2h excitations.
\cite{Hudson82a,Schulten72a,Ramasesha84a,Tavan87a,Baeriswyl92a,Schreiber08a,
Silva10a,Starcke06a,Knippenberg12a,Krauter13a,Silva08b,Schmidt12a,Tavan79a,Marian08a,Knippenberg10a}.
Since 1e$-$1h excitations, however, dominate the 2$^{1}$A$_{g}^{-}$ wave function in
the 2D compounds, especially in HBC, the analogy with the above VB qualitative picture suggests that the 
2$^{1}$A$_{g}^{-}$ in the PAH is presumably more ionic than in 1D. While investigation of the ionicity of excited states of
such large molecules is a formidable problem, we note that within VB theory large covalent character (small ionicity) of the
2$^{1}$A$_{g}^{-}$ necessarily requires large covalent character of the ground state, which can be investigated even for large
molecules within the PPP model using a variety of approximate methods such as quantum Monte Carlo. We intend to investigate
the relative covalent characters of polyenes and the present PAH molecules in the future using one such approximate method, the Path
Integral Renormalization Group, used previously by one of us in a different context. \cite{Dayal12a}
\par To conclude, we have implemented very high level correlated-electron
calculations for the one- and two-photon energy states and absorption
spectra of three PAH molecules: coronene, HBC and circumcoronene,
all possessing $D_{6h}$ point group symmetry. For coronene and HBC
we have also performed precise one- and two-photon optical absorption
experiments. The calculated one-photon energies of the 1$^{1}$B$_{2u}^{+}$
and 1$^{1}$B$_{3u}^{+}$ give excellent fits to the experimental
transition energies of these states in both coronene and HBC. For
the first time, two-photon states below the lowest optically allowed
one-photon state are demonstrated in 2D PAHs, in complete contradiction
to the predictions of the one-electron H\"uckel model. Theoretical determination
of subgap two-photon states can be achieved only by taking into account
electron correlation effects that are an order higher than the 1e$-$1h
excitonic interactions.\cite{Hudson82a,Schulten72a,Ramasesha84a,Tavan87a,Baeriswyl92a,Schreiber08a,
Silva10a,Starcke06a,Knippenberg12a,Krauter13a,Silva08b,Schmidt12a,Tavan79a,Marian08a,Knippenberg10a}. Our theoretical and experimental works, taken
together, indicate that while the consequences of e$-$e interactions
in the PAHs are weaker than in polyenes, these effects are nevertheless
significant. In particular, correlation effects in PAHs appear to
be stronger than in the phenyl group-containing $\pi$-conjugated
quasi-1D polymers poly-(paraphenylene) and poly-(paraphenylenevinylene),
in which the 2$^{1}$A$_{g}^{-}$ occurs \textit{above} the 1$^{1}$B$_{u}^{+}$
state. \cite{Soos93a} 
\par In the future we are planning to perform similar theoretical as well as
experimental investigations of PAH molecules larger than the ones
considered here. Such analyses can potentially open up an effective
means to determine the effective short-range Hubbard $U$ in graphene,
which as we pointed out in the Introduction, remains controversial. 
\begin{figure}
\includegraphics[width=3.5in]{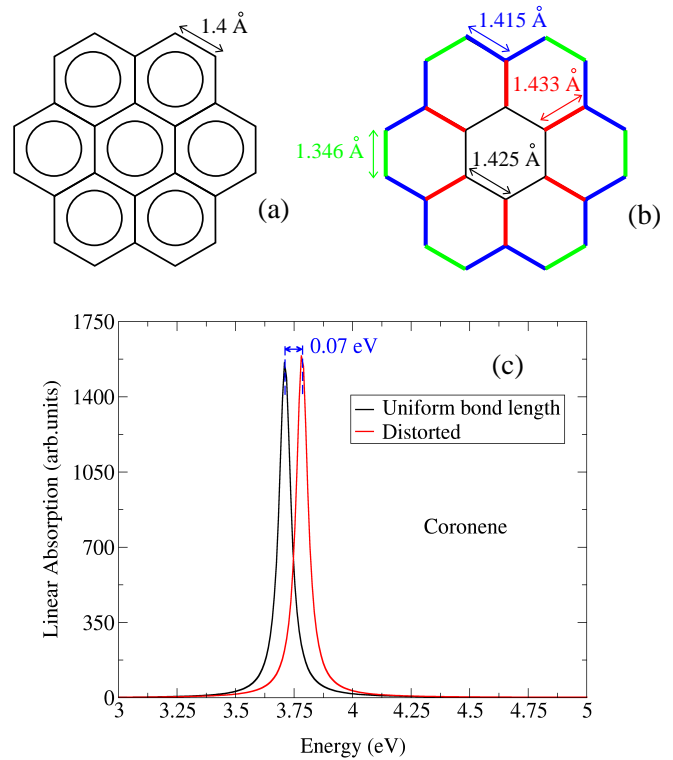}
\caption{(a) Coronene molecule with uniform bond lengths of $1.4$ $\mathring{\textrm{A}}$ assigned to all C$-$C bonds. 
(b) Actual bond lengths of coronene, from ref~\onlinecite{Fawcett65}. Note that $D_{6h}$ symmetry is maintained. (c) Ground state linear 
absorption corresponding to (a) and (b), performed using the single-CI approximation.}
\label{f4}
\end{figure}
\section*{$\blacksquare$ APPENDIX}
\subsection*{Consequences of Deviations from Idealized Geometry: The Case of Coronene} 
\label{appendix}
\par In our theoretical analyses of e$-$e correlation effects, we have used idealized geometries with uniform C$-$C bond lengths 
of $1.4$ $\mathring{\textrm{A}}$ and hopping integrals $t=2.4$ eV. In reality, the C$-$C bonds deviate from this idealized length, and
these deviations can in principle influence our overall results. Deviations from the idealized geometry are, however, expected to be small
for the aromatic molecules in question. The experimental bond lengths are available only for coronene. Here we demonstrate explicitly
for this case that the expected changes in energies due to geometry modifications are negligible, and in no way affects our main
conclusion.  
\par In Figure~\ref{f4}a,b we have shown the coronene molecule with its idealized uniform bond lengths, and with the actual bond lengths, \cite{Fawcett65}
 respectively. We calculated all hopping integrals from the known bond length$-$hopping integral
relationship, $t_{C-C}=-2.4+3.20(R_{C-C}-1.4)$, \cite{Ducasse82} where $R_{C-C}$ is the C$-$C bond length in \AA. The calculated hopping integrals
in descending order are $2.57$, $2.35$, $2.32$, and $2.29$ eV, respectively. The $V_{ij}$ were calculated as defined in eq \ref{PPP_Ham} within the text. Figure~\ref{f4}c shows the ground state linear absorption within the single-CI for the geometries of 
Figure~\ref{f4}a,b. Deviation from equal bond lengths blue shifts the main absorption band very weakly (the energy of the optical
state is slightly smaller than what is obtained within the MRSDCI, but this is of no concern here). 
\par As mentioned above, experimental bond lengths are currently not available for HBC. Based on our results for coronene, we believe, however, that
it is safe to assume that any deviations in energies due to weak distortions from idealized geometries will not affect our overall
qualitative conclusion about the reversed excited state ordering, which cannot be obtained without explicit inclusion of strong e$-$e
interactions, and which is the core theme of our work. 
\section*{$\blacksquare$ AUTHOR INFORMATION}
\noindent\textbf{Corresponding Author} \\
 {$^{*}$}E-mail: karana@physics.arizona.edu; Phone: +1-520-621-6798; Fax: +1-520-621-4721. \\
 \textbf{Notes} \\
 The authors declare no competing financial interest. 
\section*{$\blacksquare$ ACKNOWLEDGMENTS}
This work was supported by NSF Grant No. CHE-1151475. A.R. acknowledges
the support from the DoD SMART program. A.Sh. and S.M. acknowledge support
from the Indo-US Science and Technology Forum Award 37-2012/2013-14. 

\begin{thebibliography}{57}
%
\bibitem{Hudson82a} Hudson, B. S.; Kohler, B. E.; Schulten, K.
 Linear Polyene Electronic-Structure and Potential. 
\textit{Excited States.} 
\textbf{1982,} \textit{6,} 1$-$95. 
%
\bibitem{Christensen08a} Christensen, R. L.; Galinato, M. G. I.; Chu, E. F.; Howard, J. N.; 
Broene, R. D.; Frank, H. A.
Energies of Low-Lying Excited States of Linear Polyenes.
\textit{J.\ Phys.\ Chem.\ A} \textbf{2008,} 
\textit{112,} 12629$-$12636. 
%
\bibitem{Schulten72a} Schulten, K.; Karplus, M. 
On the Origin of a Low-Lying Forbidden Transition in Polyenes and Related Molecules. 
\textit{Chem.\ Phys.\ Lett.} \textbf{1972,} \textit{14,} 305$-$309. 
%
\bibitem{Ramasesha84a} Ramasesha, S.; Soos, Z. G. 
Correlated States in Linear Polyenes, Radicals, and Ions: Exact PPP Transition Moments and Spin Densities. 
\textit{J.\ Chem.\ Phys.}
\textbf{1984,} \textit{80,} 3278$-$3287. 
%
\bibitem{Tavan87a} Tavan, P.; Schulten, K. 
Electronic Excitations in Finite and Infinite Polyenes.
\textit{Phys.\ Rev.\ B}
\textbf{1987,} \textit{36,} 4337$-$4358. 
%
\bibitem{Baeriswyl92a} Baeriswyl, D.; Campbell, D. K.; Mazumdar, S. 
Conjugated Conducting Polymers; Kiess, H., Ed.; Springer: Berlin, 1992. 
%
\bibitem{Barcza13a} Barcza, G.; Barford, W.; Gebhard, F.; Legeza, O.
Excited States in Polydiacetylene Matrix Renormalization Group Study.
\textit{Phys.\ Rev.\ B} 
\textbf{2013,} \textit{87,} 245116$-$245131. 
%
\bibitem{Pariser53a} Pariser, R.; Parr, R. G. 
A Semi-Empirical Theory of the Electronic Spectra and Electronic Structure of Complex Unsaturated Molecules. II. 
\textit{J.\ Chem.\ Phys.}
\textbf{1953,} \textit{21,} 767$-$776. 
%
\bibitem{Pople53a} Pople, J. A.
 Electron Interaction in Unsaturated Hydrocarbons.
\textit{Trans.\ Faraday\ Soc.}
\textbf{1953,} \textit{49,} 1375$-$1385. 
%
\bibitem{Schreiber08a} Schreiber, M.; Silva-Junior, M. R.; Sauer, S. P. A.; Thiel, W. 
Benchmarks for Electronically Excited States: CASPT2, CC2, CCSD, and CC3.
\textit{J.\ Chem.\ Phys.} 
\textbf{2008,} \textit{128,} 134110$-$134134.
%
\bibitem{Silva10a} Silva-Junior, M. R.; Schreiber, M.; Sauer, S. P. A.; Thiel, W. 
Benchmarks for Electronically Excited States: Basis Set Effects on CASPT2 Results.
\textit{J.\ Chem.\ Phys.} 
\textbf{2010,} \textit{133,} 174318$-$174330.
%
\bibitem{Starcke06a} Starcke, J. H.; Wormit, M.; Schirmer, J.; Dreuw, A.
How Much Double Excitation Character Do the Lowest Excited States of Linear Polyenes Have?
\textit{Chem.\ Phys.} 
\textbf{2006,} \textit{329,} 39$-$49. 
%
\bibitem{Knippenberg12a} Knippenberg, S.; Rehn, D. R.; Wormit, M.; Starcke, J. H.; Rusakova, 
I. L.; Trofimov, A. B.; Dreuw, A. 
Calculations of Nonlinear Response Properties Using the Intermediate State Representation and 
the Algebraic-Diagrammatic Construction Polarization Propagator Approach: Two-Photon Absorption Spectra. 
\textit{J.\ Chem.\ Phys.}
\textbf{2012,} \textit{136,} 064107$-$064121. 
%
\bibitem{Krauter13a} Krauter, C. M.; Pernpointner, M.; Dreuw, A.
Application of the Scaled-Opposite-Spin Approximation to the Algebraic Diagrammatic Construction Schemes of Second Order. 
\textit{J.\ Chem.\ Phys.} 
\textbf{2013}, \textit{138,} 044107$-$044118.
%
\bibitem{Silva08b} Silva-Junior, M. R.; Schreiber, M.; Sauer, S. P. A.; Thiel, W. 
Benchmarks for Electronically Excited States: Time-Dependent Density Functional Theory and Density 
Functional Theory Based Multireference Configuration Interaction.  
\textit{J.\ Chem.\ Phys.} 
\textbf{2008}, \textit{129,} 104103$-$104116.
%
\bibitem{Schmidt12a} Schmidt, M.; Tavan, P. 
Electronic Excitations in Long Polyenes Revisited. 
\textit{J.\ Chem.\ Phys.} 
\textbf{2012}, \textit{136,} 124309$-$124321.
%
\bibitem{Tavan79a} Tavan, P.; Schulten, K. 
Correlation Effects in the Spectra of Polyacenes. 
\textit{J.\ Chem.\ Phys.}
\textbf{1979,} \textit{70,} 5415$-$5421. 
%
\bibitem{Raghu02a} Raghu, C.; Anusooya Pati, Y.; Ramasesha, S. 
Density-Matrix Renormalization-Group Study of Low-Lying Excitations of Polyacene within a Pariser$-$Parr$-$Pople Model.
\textit{Phys.\ Rev.\ B}
\textbf{2002,} \textit{66,} 035116$-$035126. 
%
\bibitem{Marian08a} Marian, C. M.; Gilka, N. 
Performance of the Density Functional Theory/Multireference Configuration Interaction Method on Electronic Excitation of 
Extended $\pi$-Systems.
\textit{J.\ Chem.\ Theory.\ Comput.}
\textbf{2008,} \textit{4,} 1501$-$1515. 
%
\bibitem{Knippenberg10a} Knippenberg, S.; Starcke, J. H.; Wormit M.; Dreuw, A. 
The Low-Lying Excited States of Neutral Polyacenes and Their Radical Cations: A Quantum Chemical Study Employing 
the Algebraic Diagrammatic Construction Scheme of Second Order.
\textit{Mol.\ Phys.} 
\textbf{2010,} \textit{108,} 2801$-$2813.
%
\bibitem{Geim07a} Geim, A. K.; Novoselov, S. 
The Rise of Graphene.
\textit{Nat.\ Mat.}
\textbf{2007,} \textit{6,} 183$-$191. 
%
\bibitem{Neto09a} Castro Neto, A. H.; Guinea, F.; Peres, N. M. R.; Novoselov, K. S.; Geim, A. K. 
The Electronic Properties of Graphene.
\textit{Rev.\ Mod.\ Phys.} 
\textbf{2009,} \textit{81,} 109$-$162. 
%
\bibitem{Elias11a} Elias, D. C.; Gorbachev, R. V.; Mayorov, A. S.;
Morozov, S. V.; Zhukov, A. A.; Blake, P.; Ponomarenko, L. A.; Grigorieva,
I. V.; Novoselov, K. S.; Guinea, F.; et al. 
Dirac Cones Reshaped by Interaction Effects in Suspended Graphene.
\textit{Nat.\ Phys.}
\textbf{2011,} \textit{7,} 701$-$704. 
%
\bibitem{Kotov12a} Kotov, V. N.; Uchoa, B.; Pereira, V. M.; Guinea,
F.; Castro Neto, A. H. 
Electron$-$Electron Interactions in Graphene: Current Status and Perspectives.
\textit{Rev.\ Mod.\ Phys.} 
\textbf{2012,} \textit{84,} 1067$-$1125. 
%
\bibitem{Chandross97a} Chandross, M.; Mazumdar, S. 
Coulomb Interactions and Linear, Nonlinear, and Triplet Absorption in Poly(paraphenylenevinylene).
\textit{Phys.\ Rev.\ B}
\textbf{1997,} \textit{55,} 1497$-$1504.
%
\bibitem{Sony07a} Sony, P.; Shukla, A. 
Large-Scale Correlated Calculations of Linear Optical Absorption and Low-Lying Excited States of Polyacenes: Pariser$-$Parr$-$Pople Hamiltonian.
\textit{Phys.\ Rev.\ B}
\textbf{2007,} \textit{75,} 155208$-$155229. 
%
\bibitem{Wang06a} Wang, Z. D.; Zhao, H. B.; Mazumdar, S. 
Quantitative Calculations of the Excitonic Energy Spectra of Semiconducting Single-Walled Carbon Nanotubes within a $\pi$-Electron Model.
\textit{Phys.\ Rev.\ B}
\textbf{2006,} \textit{74,} 195406$-$195411. 
%
\bibitem{Yang09a} Yang, L.; Deslippe, J.; Park, C.-H.; Cohen, M. L.; Louie, S. G. 
Excitonic Effects on the Optical Response of Graphene and Bilayer Graphene.
\textit{Phys.\ Rev.\ Lett.} 
\textbf{2009,} \textit{103,} 186802$-$186805. 
%
\bibitem{Spataru04a} Spataru, C. D.; Ismail-Beigi, S.; Benedict, L. X.; Louie, S. G. 
Excitonic Effects and Optical Spectra of Single-Walled Carbon Nanotubes.
\textit{Phys.\ Rev.\ Lett.} 
\textbf{2004,} \textit{92,} 077402$-$077405. 
%
\bibitem{Chang04a} Chang, E.; Bussi, G.; Ruini, A.; Molinari, E.
Excitons in Carbon Nanotubes: An Ab Initio Symmetry-Based Approach.
\textit{Phys.\ Rev.\ Lett.} 
\textbf{2004,} \textit{92,} 196401$-$196404. 
%
\bibitem{Prezzi08a} Prezzi, D.; Varsano, D.; Ruini, A.; Marini, A.; Molinari, E.
Optical Properties of Graphene Nanoribbons: The Role of Many-Body Effects.
\textit{Phys.\ Rev.\ B} 
\textbf{2008,} \textit{77,} 041404$-$041407. 
%
\bibitem{Rohlfing99a} Rohlfing, M.; Louie S. G. 
Optical Excitations in Conjugated Polymers.
\textit{Phys.\ Rev.\ Lett.}
\textbf{1999,} \textit{82,} 1959$-$1962. 
%
\bibitem{Ramasesha00a} Ramasesha, S.; Pati, S. K.; Shuai, Z.; Br\'edas, J. L. 
The Density Matrix Renormalization Group Method: Application to the Low-Lying Electronic States in Conjugated Polymers.
\textit{Adv.\ Quantum \ Chem.} 
\textbf{2000,} \textit{38,} 121$-$215. 
%
\bibitem{Bursill02a} Bursill, R. J.; Barford, W. 
Large-Scale Numerical Investigation of Excited States in Poly(paraphenylene).
\textit{Phys.\ Rev.\ B}
\textbf{2002,} \textit{66,} 205112$-$205119. 
%
\bibitem{Buenker74} Buenker, R. J.; Peyerimhoff, S. D. 
Individualized Configuration Selection in CI Calculations with Subsequent Energy Extrapolation.
\textit{Theor.\ Chim.\ Acta.}
\textbf{1974,} \textit{35,} 33$-$58.
%
\bibitem{Tavan86} Tavan, P.; Schulten, K. 
The Low-Lying Electronic Excitations in Long Polyenes: A PPP-MRD-CI Study.
\textit{J.\ Chem.\ Phys.}
\textbf{1986,} \textit{85,} 6602$-$6609. 
%
\bibitem{Ghosh00} Ghosh, H.; Shukla, A.; Mazumdar, S. 
Electron-Correlation-Induced Transverse Delocalization and Longitudinal Confinement in
 Excited States of Phenyl-Substituted Polyacetylenes.
\textit{Phys.\ Rev.\ B}
\textbf{2000,} \textit{62,} 12763$-$12774. 
%
\bibitem{Shukla02a} Shukla, A. 
Correlated Theory of Triplet Photoinduced Absorption in Phenylene$-$Vinylene Chains.
\textit{Phys.\ Rev.\ B} 
\textbf{2002,} \textit{65,} 125204$-$125209. 
%
\bibitem{Shukla03a} Shukla, A.; Ghosh, H.; Mazumdar, S. 
Theory of Excited-State Absorption in Phenylene-Based $\pi$-Conjugated Polymers.
\textit{Phys.\ Rev.\ B}
\textbf{2003,} \textit{67,} 245203$-$245211. 
%
\bibitem{Shukla04a} Shukla, A. 
Theory of Two-Photon Absorption in Poly(diphenyl) Polyacetylenes.
\textit{Chem.\ Phys.} \textbf{2004,}
\textit{300,} 177$-$188. 
%
\bibitem{Shukla04b} Shukla, A. 
Theory of Nonlinear Optical Properties of Phenyl-Substituted Polyacetylenes.
\textit{Phys.\ Rev.\ B} \textbf{2004,}
\textit{69,} 165218$-$165227. 
%
\bibitem{Sony05a} Sony, P.; Shukla, A. 
Photoinduced Absorption in Disubstituted Polyacetylenes: Comparison of Theory with Experiments.
\textit{Phys.\ Rev.\ B}
\textbf{2005,} \textit{71,} 165204$-$165208. 
%
\bibitem{Aryanpour11a} Aryanpour, K.; Sheng, C. X.; Olejnik, E.; Pandit, B.; Psiachos, D.; Mazumdar, S.; Vardeny, Z. V. 
Evidence for Excimer Photoexcitations in an Ordered $\pi$-Conjugated Polymer Film.
\textit{Phys.\ Rev.\ B}
\textbf{2011,} \textit{83,} 155124$-$155128. 
%
\bibitem{Boyd92a} Boyd, R. W. 
Nonlinear Optics; Academic Press, Inc.: San Diego, CA, 1992. 
%
\bibitem{Heflin88a} Heflin, J. R.; Wong, K. Y.; Zamani-Khamiri, O.; Garito, A. F.
Nonlinear Optical Properties of Linear Chains and Electron-Correlation Effects.
\textit{Phys.\ Rev.\ B} 
\textbf{1988,} \textit{38,} 1573$-$1576. 
% 
\bibitem{McWilliams91a} McWilliams, P. C. M.; Hayden, G. W.; Soos, Z. G. 
Theory of Even-Parity States and Two-Photon Spectra of Conjugated Polymers.
\textit{Phys.\ Rev.\ B} 
\textbf{1991,} \textit{43,} 9777$-$9791. 
%
\bibitem{FGuo94a} Mazumdar, S.; Guo, F. 
Intensities of Two-Photon Absorptions to Low-Lying Even-Parity States in Linear-Chain Conjugated Polymers.
\textit{Phys.\ Rev.\ B}
\textbf{1994,} \textit{49,} 10102$-$10112. 
%
\bibitem{Allan84a} Allan, M.; Neuhaus L.; Haselbach E. 
(all-E)-1,3,5,7-Octatetraene: Electron-Energy-Loss and Electron-Transmission Spectra.
\textit{Helv.\ Chim.\ Acta}
\textbf{1984,} \textit{67,} 1776$-$1782. 
%
\bibitem{Abouaf09a} Abouaf, R.; D\'{i}az-Tendero S. 
Electron Energy Loss Spectroscopy and Anion Formation in Gas Phase Coronene.
\textit{Phys.\ Chem.\ Chem.\ Phys.}
\textbf{2009,} \textit{11,} 5686$-$5694. 
%
\bibitem{Guclu10a} G\"u\c cl\"u, A. D.; Potasz, P.; Hawrylak, P. 
Excitonic Absorption in Gate-Controlled Graphene Quantum Dots.
\textit{Phys.\ Rev.\ B}
\textbf{2010,} \textit{82,} 155445$-$155449. 
%
\bibitem{Rathore03a} Rathore, R.; Burns, C. L. 
A Practical One-Pot Synthesis of Soluble Hexa-Peri-Hexabenzocoronene and Isolation of Its Cation-Radical Salt.
\textit{J.\ Org.\ Chem.}
\textbf{2003,} \textit{68,} 4071$-$4074. 
%
\bibitem{SheikBahae90a} Sheik-Bahae, M.; Said, A. A.; Wei, T.-H.; Hagan, D. J.; Van Stryland, E. W. 
Sensitive Measurement of Optical Nonlinearities Using a Single Beam.
\textit{IEEE.\ J.\ Quantum Electron.}
\textbf{1990,} \textit{26,} 760$-$769. 
%
\bibitem{Menard07a} M\'enard, J. M.; Betz, M.; Sigal, I.; van Driel, H. M.
Single-Beam Differential Z-Scan Technique.
\textit{Appl.\ Opt.} 
\textbf{2007,} \textit{46,} 2119$-$2122. 
%
\bibitem{Dayal12a} Dayal, S.; Clay, R. T.; Mazumdar, S. 
Absence of Long-Range Superconducting Correlations in the Frustrated Half-Filled-Band Hubbard Model.
\textit{Phys.\ Rev.\ B} 
\textbf{2012,} \textit{85,} 165141$-$165148. 
%
\bibitem{Soos93a} Soos, Z. G.; Ramasesha, S.; Galv\~ao, D. S. 
Band to Correlated Crossover in Alternating Hubbard and Pariser-Parr-Pople Chains: Nature of the Lowest 
Singlet Excitation of Conjugated Polymers.
\textit{Phys.\ Rev.\ Lett.}
\textbf{1993,} \textit{71,} 1609$-$1612. 
%
\bibitem{Fawcett65} Fawcett, J. K. 
The determination and refinement of the molecular structures of some organic compounds; 
http://hdl.handle.net/2429/38382, 1965.
%
\bibitem{Ducasse82} Ducasse, L. R.; Miller, T. E.; Soos, Z. G. 
Correlated States in Finite Polyenes: Exact PPP Results.
\textit{J.\ Chem.\ Phys.} 
\textbf{1982,} \textit{76,} 4094$-$4104.
%
\end{thebibliography}
\end{document}